\documentclass[
 reprint,
superscriptaddress,
groupedaddress,
frontmatterverbose, 
 amsmath,amssymb,
 prl,
floatfix,
]{revtex4-2}

\usepackage{graphicx}
\usepackage{dcolumn}
\usepackage{bm}
\usepackage{hyperref}
\hypersetup{
  colorlinks = true,
  allcolors = black
}
\usepackage{xcolor}

\begin{document}

\title{Validation of machine-learned interatomic potentials via temperature-dependent electron thermal diffuse scattering}

\author{Dennis S.~Kim}
\email{dennis.s.kim@icloud.com}
\affiliation{
Department of Materials Science \& Engineering, Massachusetts Institute of Technology, Cambridge, MA 02139
}
\author{Michael Xu}
\affiliation{ 
Department of Materials Science \& Engineering, Massachusetts Institute of Technology, Cambridge, MA 02139
}
\author{James M.~LeBeau}%
\email{lebeau@mit.edu}
\affiliation{ 
Department of Materials Science \& Engineering, Massachusetts Institute of Technology, Cambridge, MA 02139
}

\date{\today}

\begin{abstract}
Machine-learned interatomic potentials (MLIPs) show promise in accurately describing the physical properties of materials, but there is a need for a higher throughput method of validation. Here, we demonstrate using that MLIPs and molecular dynamics can accurately capture the potential energy landscape and lattice dynamics that are needed to describe electron thermal diffuse scattering. Using SrTiO$_3$ as a test-bed at cryogenic and room temperatures, we compare electron thermal diffuse scattering simulations using different approximations to incorporate thermal motion. Only when the simulations are based on quantum mechanically accurate MLIPs in combination with path-integral molecular dynamics that include nuclear quantum effects, there is excellent agreement with experiment\end{abstract}

\maketitle
Thermally excited atomic motion governs many thermophysical properties of solid-state materials and have macroscopic consequences on phase transitions, physical, chemical, electronic, and optoelectronic material properties~\cite{Zhong1996-cx,He2020-ki,Lofgren2016-mv,Kim2018-zx}. Modeling this correlated atomic motion is often considered within the harmonic approximation (first-order terms of the Taylor expansion) of the crystal potential energy surface.
At finite temperatures, however, many solids behave anharmonically, in part, as a consequence of phonon-phonon interactions.
Further, the higher-order terms describing the energy surface lead to finite phonon lifetimes or energy broadening, which are essential to describe thermal transport properties~\cite{Peierls1929-fo}. In addition, nuclear quantum effects are often neglected, which includes the zero-point energy and atomic tunneling~\cite{Markland2018-mb}. Machine-learned interatomic potentials (MLIPs), which are used as classical molecular dynamic potentials with local descriptors, have been shown to be capable of accurately mapping the potential energy surface of a system and offer the prospect to circumvent the size constraints in first-principles calculations \cite{Zuo2020-gj,Thompson2015-eq,Wood2018-hx}.  
These MLIPs are, however, typically tested against simulated or experimental phonon dispersions, i.e.~the frequency-wavevector relations. Assessing phonon energies and lifetimes at spatial scales relevant for modern nanoscale applications have, however, proven difficult~\cite{Gadre2022-uh,Tian2021-vq}. 


Infrared, Raman, neutron, X-ray, and electron scattering methods have made significant progress in measuring phonon behavior in bulk materials, each with its own limiation. 
Infrared and Raman scattering techniques, for instance, are typically limited to only symmetry-allowed modes, while neutron and X-ray methods generally require bulk samples and instrumentation only available at national facilities. 
Meanwhile, recent advances in electron energy loss spectroscopy in scanning transmission electron microscopy (STEM) enable direct measurement of vibrational properties with nanometer resolution~\cite{Gadre2022-uh}, but require specialized instrumentation and are most sensitive to higher-energy phonon modes~\cite{Hachtel2018-cl,Krivanek2014-ci}. 

Beyond the conventional methods, thermal diffuse scattering offer a possible route for high-throughput MLIP validation. Electron thermal diffuse scattering (eTDS) in parallel beam electron diffraction ~\cite{Wang2003-gq,Zuo2000-gx,Otto2021-ad,Rene_de_Cotret2022-jy}, for example, encodes the phonon dispersion relationships into non-uniform background features that arises due to partially coherent displacements of atoms away from  equilibrium positions ~\cite{Debye1914-ap,Waller1923-py,Born1942-du}. To compare theory and experiment in ultrathin single-layer materials, for example, analytical kinematical electron scattering calculations have been used as was applied in X-ray studies~\cite{Holt1999-zk}. Moreover, these prior studies have largely relied on harmonic phonons without including cubic and quartic terms in the crystal Hamiltonian (anharmonicity) and nuclear quantum effects~\cite{Muller2001-ux,Otto2021-ad,Rene_de_Cotret2022-jy}. While kinematical models are in agreement with ultrathin samples, fully quantifying analysis of eTDS has been stymied by the difficulties in data interpretation resulting from strong dynamical scattering and other inelastic scattering processes (e.g. plasmon losses). 

For comparisons with experiment, electron scattering simulations must accurately describe scattering within samples that are 10's to 100's of nm thick, beyond the limits first-principles calculations ~\cite{Muller2001-ux, Durham2022-mb,Krause2018-er}. This is largely achieved using dynamical frozen phonon multislice simulations that are the workhorse to model the effect of thermal vibrations on scattering~\cite{Cowley1957-jv,Muller2001-ux,Wang1998-xj,Rosenauer2008-se,Van_Dyck2009-hu,Forbes2011-wv}. %
Further, these frozen phonon simulations are in excellent agreement with quantum mechanically exact methods such as the quantum excitation potential method (QEP)~\cite{Forbes2010-zc}. Usually, however, only the Einstein approximation (uncorrelated thermal displacements) are considered. Although this approximation has been shown to be in excellent agreement with imaging experiments \cite{LeBeau2008-qh,Thust2009-nd}, it does not accurately reproduce the diffuse scattering details such as streaking or Brillouin zone boundary features. In contrast, frozen phonon simulations backed by the stochastic temperature-dependent effective potential method showed feasibility for predicting accurate electron diffraction patterns, but only for Si where anharmonicity and nuclear quantum effects are not the major contribution to capture lattice dynamics at room temperature ~\cite{Chen2023-mr}. As such, validating MLIPs with eTDS to understand atomic vibrations in complex systems exhibiting anharmonicity or nuclear quantum effects has so far been unexplored.

In this Letter, we show that applying state-of-the-art machine-learned interatomic potentials and molecular dynamics simulations can accurately model electron diffuse scattering experiments to provide a quantifiable approach to characterize  temperature-dependent lattice dynamics. 
Phonon anharmonicity and nuclear quantum effects are included through the use of these MLIPs with classical (ML-CMD) and path-integral molecular dynamics (ML-PIMD). 
Here, SrTiO$_3$ is chosen as a model system due to its anomalous anharmonic phonon softening with decreasing temperature, known strong electron-phonon coupling, antiferrodistortive structural phase transition, and strong quantum fluctuation dependence~\cite{Zhong1996-cx,Zhou2019-yy,Tadano2019-ju,Abramov1995-nz,Zhou2018-fi,Cowley1964-yp,Perry1967-di,Wang2000-ug,Holt2007-ai,Ravy2007-nh,Neumann1995-fp,Rutt1997-tg,Hunnefeld2002-te}. 
Frozen phonon simulations backed by MLIPs based thermal displacements are compared against  parallel beam diffraction patterns  captured high-dynamic range methods. Only after accounting for the key contributions to thermal displacements, the calculations are shown to be in excellent agreement with the experiments. This method is general and can complement other methods to understand changes in atomic motion caused by many-body interactions, defects, and/or finite-size effects in solid materials.


SrTiO$_3$ single crystals were acquired from MTI Corporation (Richmond, CA)  and  thinned to electron transparency by conventional wedge polishing followed by argon ion milling using a Fischione (Export, PA) Model 1051 mill.  
A Thermo Fisher Scientific Themis Z 60-300 S/TEM was operated at 200\,kV to collect energy-filtered parallel beam diffraction patterns. Electrons with energy losses greater than 2~eV were removed by energy filtering with a Gatan Continuum 1066 HR image filter equipped with a scintillator-based CMOS camera. The dynamic range of the camera was enhanced by acquiring and recombining six different exposures from 50~ms to 16~s with the method outlined in Ref.~\cite{Evans2014-vb}.
Room- and low-temperature measurements were performed at roughly the same position on the sample using a Mel-Build Double tilt LN2 Atmos Defend Holder. To estimate the sample thickness, the electron energy loss, $t/\lambda$ method was used. With an inelastic mean free path of 106~nm estimated following Ref.~\cite{Egerton1987-gc}, the samples were found to be flat and have a thickness of approximately 40~nm~ in the region of interest \cite{Dennis_S_Kim_Michael_Xu_James_M_LeBeau_undated-fi}. Multislice simulations were conducted with the abTEM Python package~\cite{Madsen2021-yt,Kirkland2010-bv} with conditions matching the experiment. All simulated supercells were 8.19 nm$~\times$~8.27 nm$~\times~32.54~$ nm. Using the known sample thickness, we then used multislice simulations that account for different theoretical approximations of atomic vibrations, including the Einstein approximation, harmonic approximated displacements, and molecular dynamics-based anharmonicity included displacements with and without nuclear quantum effects. In each case, the reported simulations are the result of an incoherent averaging of 100 different thermal configurations.

\begin{figure}
    \centering
    \includegraphics[width=0.45\textwidth]{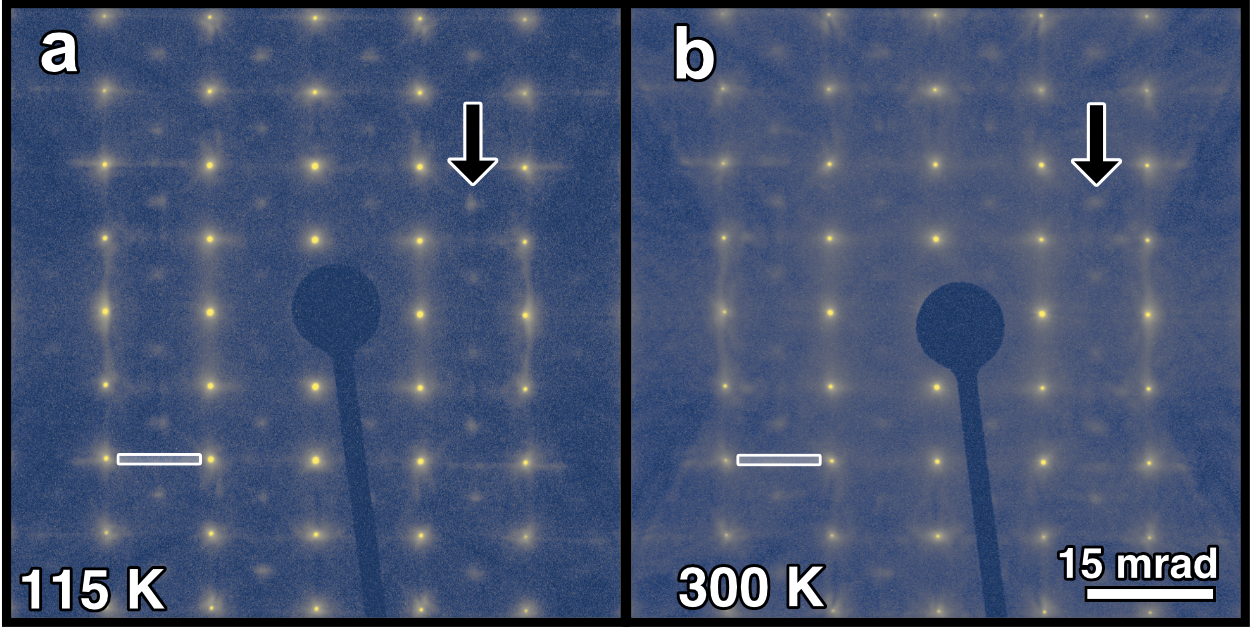}
    \caption{High-dynamic range spot diffraction patterns including thermal diffuse scattering (TDS) of [110] SrTiO$_3$ at 115~K (a) and 300~K (b). Black arrows indicates the BZ R-point intensity and white box highlights the linear thermal diffuse features.}
    \label{fig:exp}
\end{figure}

High dynamic range parallel beam diffraction patterns of cubic SrTiO$_3$ along $\left[110\right]$ at 115~K (a) and 300~K (b) are shown in Figure~\ref{fig:exp}. 
The linear features between Bragg reflections, such as indicated by the white box, are seen throughout the patterns and are the result of both thermal diffuse scattering and Kikuchi bands.  
An example of thermal diffuse linear features or streaks are enclosed with a white box in Fig.~\ref{fig:exp}. 
In addition to the strong Bragg reflections (brightest features) and Kikuchi bands, the intensity also peaks at positions in between reflections for both temperatures, as black arrows in Fig.~\ref{fig:exp}. 
Notably, these peaks occur at the R-point zone-boundary points, which should not be observed given the cubic \textit{Pm$\bar{3}$m} crystal structure.
Moreover, as a function of temperature, these features are considerably sharper (narrower) at 115 K compared to 300~K.
Specifically, the average full-width at half maximum of the R-point feature changed on average by 15.8\% between 300~K and 115~K.
This is consistent with the trend in previous X-ray scattering experiments~\cite{Holt2007-ai}, and has been attributed to the temperature-dependent softening of the R-point phonon mode~\cite{Shirane1969-pk,Andrews1986-if,Nelmes1988-yu,Holt2007-ai,He2020-ki,Ravy2007-nh}.

To test the accuracy of and to validate the machine-learned interatomic potentials for the highly anharmonic SrTiO$_3$, we examined the lattice dynamics simulations using the stochastic temperature-dependent effective potential method (sTDEP) comparing \textit{ab initio} and quadratic spectral neighbor analysis potentials (qSNAP) potentials~\cite{Hellman2013-fu,Hellman2013-td,Zuo2020-gj,Thompson2015-eq,Wood2018-hx}. Although multiple MLIPs can accurately depict the potential energy surface of SrTiO$_3$, due to computational speeds, here we focus on the qSNAP potential (Supplemental Materials)~\cite{Dennis_S_Kim_Michael_Xu_James_M_LeBeau_undated-fi,Zuo2020-gj,Thompson2015-eq,Wood2018-hx,He2022-qz}. First-principle simulations were performed using \textsc{Quantum Espresso} with a revised Perdew-Burke-Ernzerhof generalized gradient approximation (PBEsol) functional to calculate exchange-correlation energies~\cite{Giannozzi2009-ru,Giannozzi2017-xu,Perdew2008-mg}.
We also calculated 0~K Harmonic phonon energies for comparison.
The \textsc{sTDEP} method fits the effective second- and higher-order force constants of the crystal Hamiltonian (Eq.~S1) from forces calculated from thermally displaced atoms in supercells (Eq.~S2).
The anharmonic ($H_a$) components of the Hamiltonian include higher-order terms where the third- and fourth-order terms and force constants are shown in Eq.~S1.
The anharmonic correction to phonon self-energy ($\Sigma$) determined by the third- and fourth-order terms in Eq.~S1 has real ($\Delta$) and imaginary ($i\Gamma$) parts and gives rise to thermal shifts and energy broadening, respectively.
The atomic displacements were based on stochastic samplings of the canonical ensemble and followed the Planck's distribution and included nuclear quantum effects~\cite{Shulumba2017-si,Kim2018-zx}.
For more information on computational methods see the Supplemental Materials~\cite{Dennis_S_Kim_Michael_Xu_James_M_LeBeau_undated-fi}. 

\begin{figure}
    \centering
    \includegraphics[width = 3.1in]{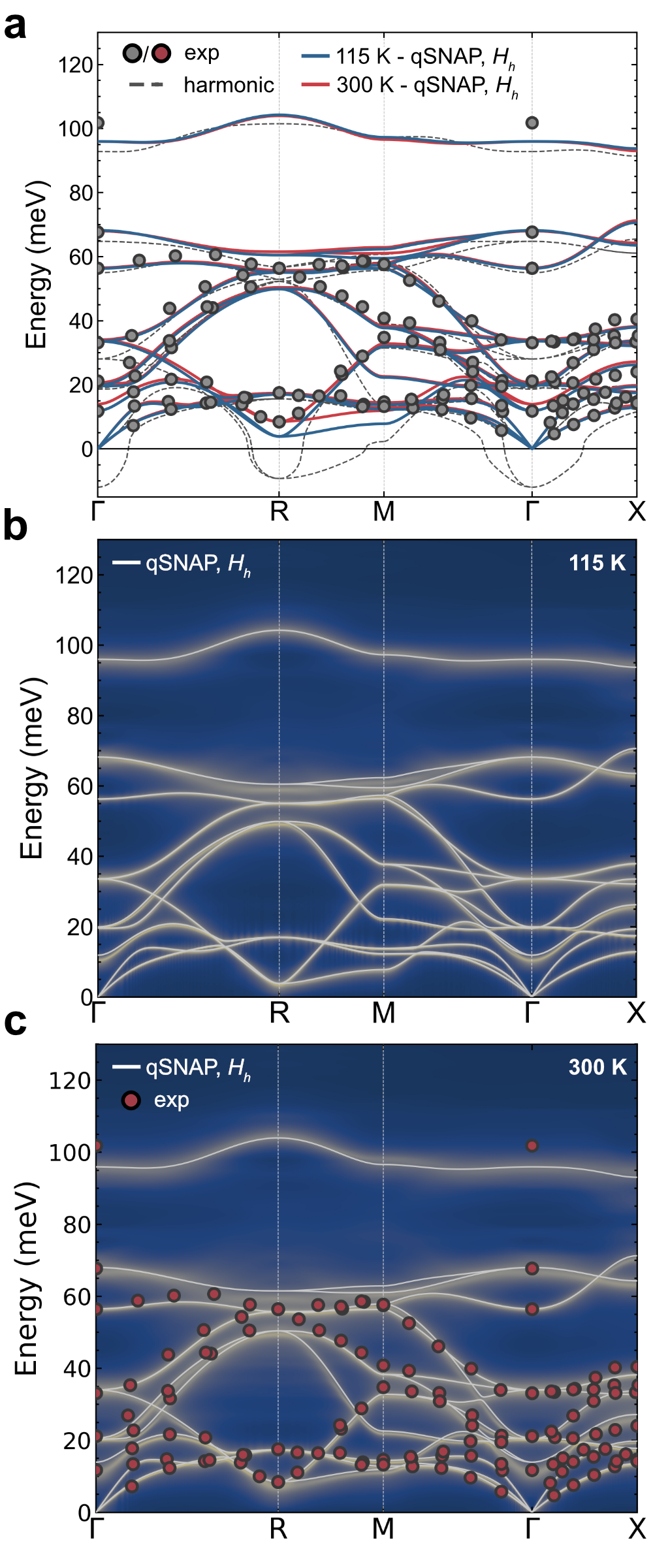}
    \caption{Phonon dispersions along high-symmetry directions of SrTiO$_3$. Phonon energies were calculated using the temperature-dependent effective potential method (TDEP) using the qSNAP machine-learned interatomic potential (solid line) at 300~K (red) and 115~K (blue). Inelastic neutron and Raman scattering experiments are shown as gray circle markers\,\cite{Stirling1972-di,Servoin1980-vw}. Phonon spectral functions were calculated using Supplemental Eq.~3 at 115~K (b) and 300~K (c). Harmonic phonon dispersions at 0~K are shown as a grey dashed line in a-c. Inelastic neutron and Raman scattering experiments are shown as gray circle markers in (a) and maroon circle markers in (c)~\cite{Stirling1972-di,Servoin1980-vw}}
    \label{fig:dispersions}
\end{figure}

Unlike the harmonic approximated phonon band structure, we find that the machine-learned interatomic potentials reliably reproduce the potential energy surface by comparing the energies, forces, phonon dispersions, and phonon spectral functions (Supplemental Materials)~\cite{Dennis_S_Kim_Michael_Xu_James_M_LeBeau_undated-fi,He2022-qz}. 
As shown previously, cubic-SrTiO$_3$ is anharmonically stabilized as evidenced by the imaginary frequencies in the 0~K harmonic approximation (grey dashed line) compared to predicted stable structures in anharmonically renormalized harmonic dispersions at 115~K (blue solid line) and 300~K (red solid line) in Fig.~\ref{fig:dispersions}a. Moreover, the qSNAP potential can reproduce the effects of phonon anharmonicity beyond the renormalized harmonic frequencies as seen in the spectral functions in Fig.~\ref{fig:dispersions}b-c and Supplemental Fig.~S1 (Supplemental Materials)~\cite{Dennis_S_Kim_Michael_Xu_James_M_LeBeau_undated-fi}. Also, strong phonon thermal broadening is apparent in high-energy optical modes, and especially in soft-phonon modes near the R-point and the low energy $\Gamma$-point modes (Fig.~\ref{fig:dispersions}c).

\begin{figure*}
    \centering
    \includegraphics[width=0.85\textwidth]{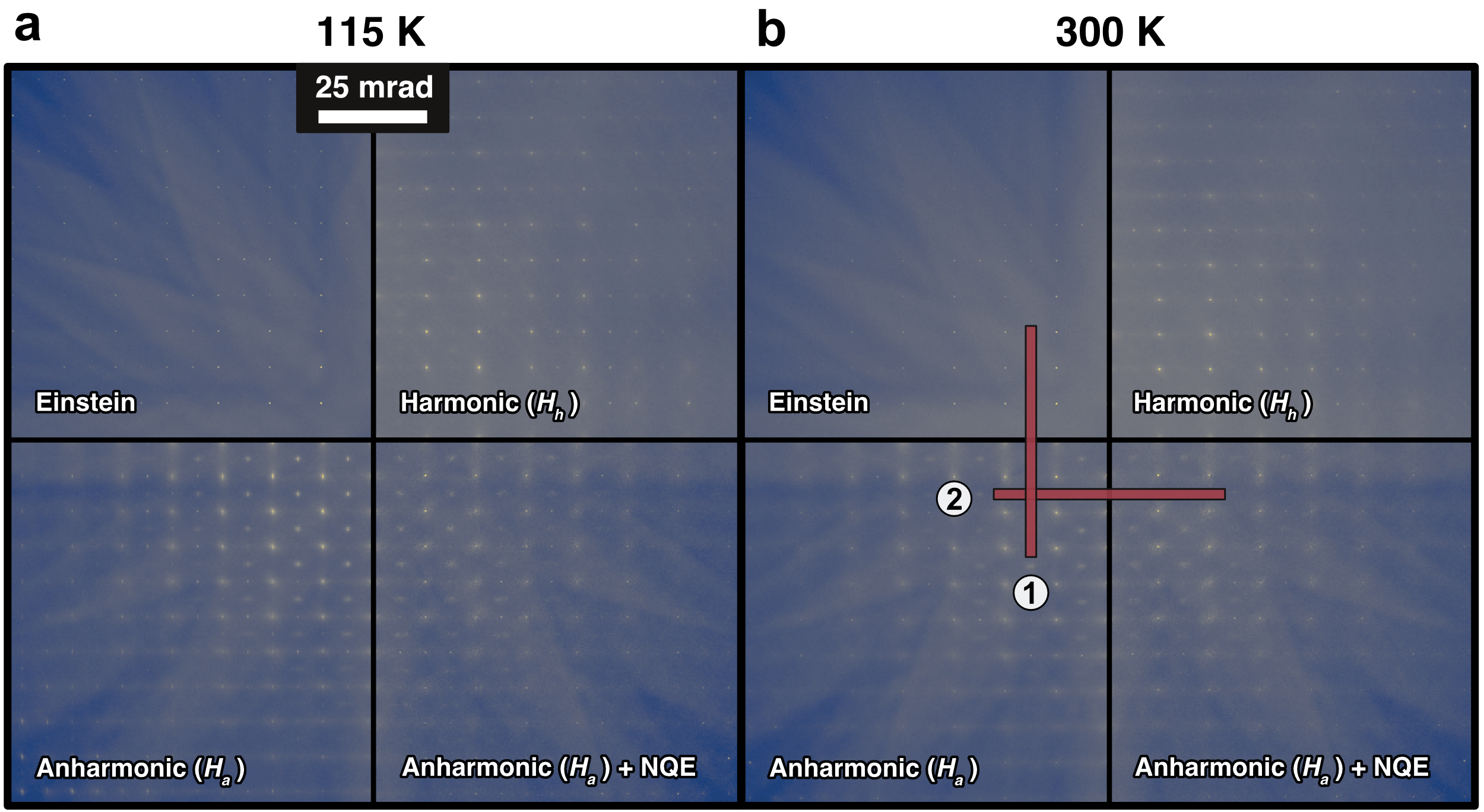}
    \caption{Frozen phonon dynamical simulations using machine-learned interatomic potentials. Multislice simulations of parallel beam electron diffraction patterns of 32.5~nm thick SrTiO$_3$ [110] slabs with a zone-axis at 115~K (a) and 300~K (b). Einstein (top-left), harmonic ($H_2$) approximated displacements using finite-temperature stochastic displacements using renormalized phonon dispersions  (top-right), anharmonicity included displacements from machine-learned interatomic potential based molecular dynamics (bottom-left), and anharmonicity and nuclear quantum effects included displacements (bottom-right) are shown for both temperatures. Line profiles of thermal diffuse scattering along the respective line in (1) and (2) are compared with experiments in Fig.~\ref{fig:dyn_fit}.}
    \label{fig:dynamical}
\end{figure*}

To understand the effects of different approximations on the thermal atomic displacements, we performed  frozen-phonon multislice simulations for each.
Equipped with an MLIP that accurately describes the finite-temperature potential energy surface and phonon dispersions from first principles, molecular dynamic simulations are performed to  include the atomic interactions of higher-order terms in the Hamiltonian of simulated structures with thicknesses matching experiment.
Supplemental Fig.~S2 shows a typical cell on the order of nanometers that is typically used for first-principles calculations, where machine-learned interatomic potentials allow for similar accuracies of forces on atoms with the computational cost that allows for large simulation sizes that can approach tens to hundreds of nanometers in width and thickness~\cite{Dennis_S_Kim_Michael_Xu_James_M_LeBeau_undated-fi}. 
This reduction in computational cost also allows for the addition of computing nuclear quantum effects in atomic dynamics, as Supplemental Fig.~S2.c shows a cut of SrTiO$_3$ super-imposed with 16 beads, where each represents a classical particle that makes up the quantum mechanical particles used in path-integral molecular dynamics simulations~\cite{Markland2018-mb}.


First, consider the Einstein approximation, which accounts for the correct magnitude of thermal displacements by assuming that atoms behave as independent harmonic oscillators. Although this approximation works well in simulating Kikuchi bands \cite{Loane1991-xq,Kirkland2020-xs} and high-angle annular dark-field (HAADF) imaging \cite{LeBeau2008-qh}, this simplified model does not capture the intricate collective behavior determined by the phonon dispersion relations as seen in the lack of strong intensities between Bragg reflections and at R-points in the top left panel of Fig.~\ref{fig:dynamical}a-b. 



\begin{figure}
    \centering
    \includegraphics[width=0.333\textwidth]{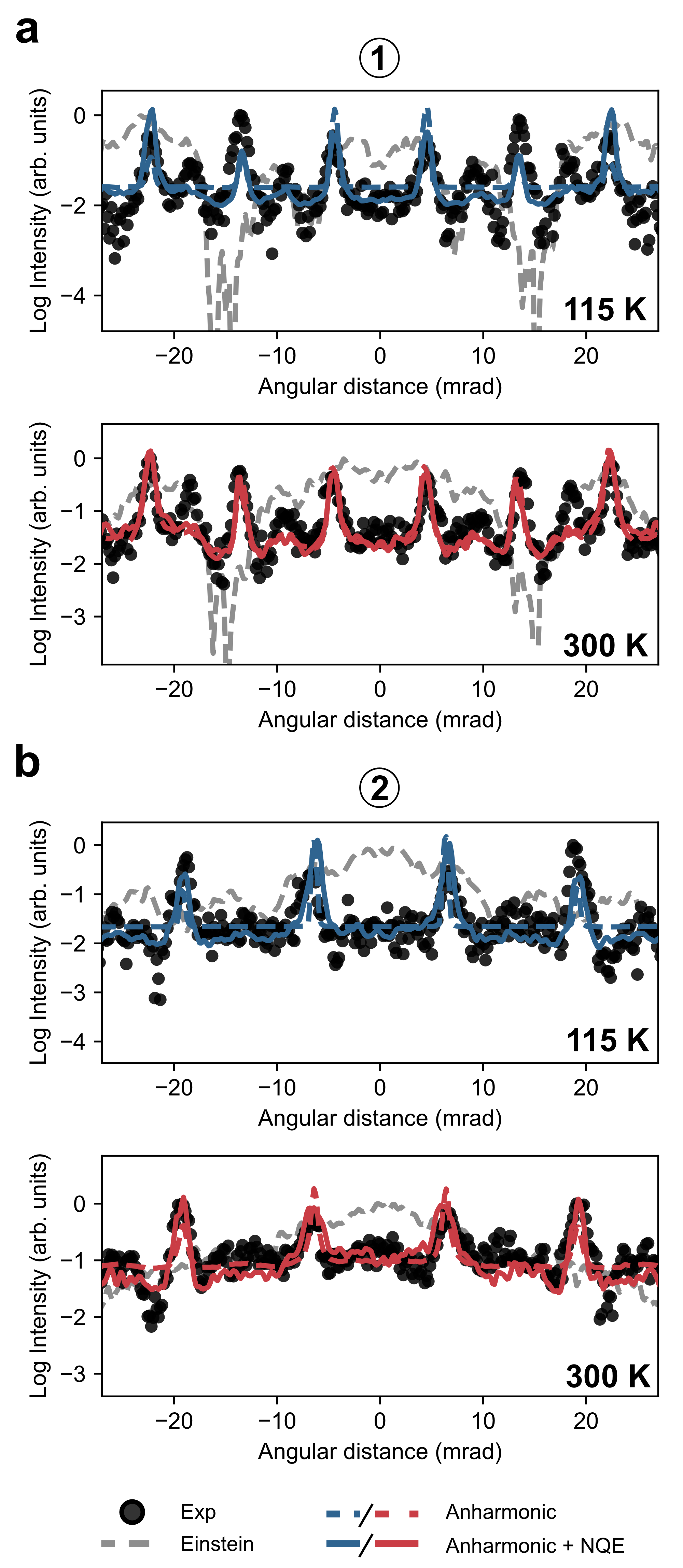}
    \caption{Line profiles comparing experimental data from Fig.~\ref{fig:exp} and dynamical simulations are shown for (1) and (2) lines of Fig.~\ref{fig:dynamical}. Experimental data are shown as black circle markers, the Einstein approximated simulations are shown as grey dashed lines, and anharmonic and anharmonic with nuclear quantum effects are shown as colored dashed and colored solid lines, respectively. Line profiles from 115~K and 300~K are shown as blue and red, respectively.}
    \label{fig:dyn_fit}
\end{figure}

The collective behavior of atomic motion can be introduced by stochastically displacing atoms based on eigenfrequencies in phonon dispersion relations~\cite{Shulumba2017-si}. 
This approximation for a canonical ensemble has proven effective when fitting interatomic force constants or using this with strongly bonded solids with small yet anharmonic thermal displacements in Si~\cite{Chen2023-mr,Kim2018-zx}.  
The scattering simulations from atomic displacements from renormalized harmonic ($H_h$) simulations reproduce streaking features (top-right of Fig.~\ref{fig:dynamical}.a and b), but do not reproduce the strong intensities at R-points represented as black arrows in the experimental data of Fig.~\ref{fig:exp}. 
This approximation includes phonon dispersion information but lacks the higher-order terms of the crystal Hamiltonian and thus any information of phonon-phonon interactions.
We observe strong intensities at these R-points in patterns only when we include phonon anharmonicities through molecular dynamics simulations  (bottom panel Fig.~\ref{fig:dyn_fit}).
We have also performed simulations for different thicknesses and found small variations in peak intensities, but all thicknesses retained the features at the R-point and do not effect the analysis (Fig.~S3 and S4).

When nuclear quantum effects, through path-integral methods, are added into the atomic motion, the thermal diffuse patterns are a better fit to experimental data.
Line scans shown in Fig.~\ref{fig:dyn_fit}a, and Fig.~\ref{fig:dyn_fit}b are from red boxes labeled (1) and (2) in Fig.~\ref{fig:dispersions}b, respectively. 
The top panels are from 115~K and the bottom from 300~K.
Einstein (grey dashed line), anharmonic ($H_a$, colored dashed line), and anharmonic ($H_a$) with nuclear quantum effects (NQE) (colored solid line) were scaled to best fit experimental data (black circle markers in Fig.~\ref{fig:dyn_fit}).  
In contrast, thermal diffuse scattered simulations, including anharmonicities with and without nuclear quantum effects, are shown as solid and dashed lines, respectively. 
From these results, the majority of the strong features are explained by including the atomic vibrations resulting from the anharmonic phonon modes. Moreover, the fit to experiment improves  further when including nuclear quantum effects, especially at cryogenic temperatures, where this contribution becomes more significant. 
The average R$^2$ for the least-square fits increase from 29.6\% to 45.6\% when nuclear quantum effects are included (Supplemental Materials Fig. S6), whereas the Einstein approximation can only fit a constant intensity. 
The full-width half-maximum of the R-point changes with temperature by 23.5\% with and 88\% without nuclear quantum effects.
In other words, the ML-CMD based simulations over estimate the sharpness of the R-point features at 115~K.
While the atomic motion in SrTiO$_3$ is known to show strong nuclear quantum effects~\cite{Zhong1996-cx}, this has been previously difficult to accurately incorporate in electron microscopy simulations described above.

In conclusion, we have measured temperature-dependent electron thermal diffuse scattering of SrTiO$_3$ and show the details of the diffuse intensity distribution can validate MLIPs.
These MLIPs with classical and path-integral molecular dynamics can account for anomalous, anharmonic phonons and nuclear quantum effects.
Moreover, the anomalous phonon soft-mode dynamics are predicted in the electron diffraction patterns and find much better agreement with simulations when phonon anharmonicity and nuclear quantum effects are included. 
For dynamical simulations, MLIPs are necessary for quantum mechanically accurate predictions due to larger-scale simulations that can account for realistic sample sizes in reasonable computational times. 
Overall, these results demonstrate that eTDS opens a pathway to reliably validate MLIPs and to quantify atomic thermal behavior in materials.

We gratefully acknowledge support for this research from the Air Force Office of Scientific Research under contract FA9550-20-0066. This work was carried out in part through the use of the MIT Characterization.nano facility. 
The authors acknowledge the MIT SuperCloud and Lincoln Laboratory Supercomputing Center for providing (HPC, database, consultation) resources that have contributed to the research results reported within this paper/report.

\bibliography{paperpile_sto_v2.bib}

\end{document}